\documentclass[epj]{svjour}

\usepackage{amsmath}
\usepackage{amssymb}
\usepackage{epsfig}
\usepackage{graphicx}
\usepackage{color}\usepackage[utf8]{inputenc}

\usepackage{amsfonts}
\usepackage{bbold}
\usepackage{mathptmx} 

\usepackage{bm}
\usepackage{xcolor}
\usepackage{fix-cm}
\usepackage[T1]{fontenc}

\usepackage[colorlinks,allcolors=blue]{hyperref}

\begin{document}

\title{Hartree-Fock-Bogoliubov study of quantum shell effects on the path to fission in $^{180}$Hg, $^{236}$U and $^{256}$Fm}

\author{R. N. Bernard\inst{1}\thanks{\emph{Present address: CEA, DES, Cadarache, F-13108 Saint Paul les Durance, France}} \and C. Simenel\inst{1,2}\thanks{\emph{cedric.simenel@anu.edu.au}} \and G. Blanchon\inst{3,4}}
\institute{Department of Fundamental and Theoretical Physics, Research School of Physics, The Australian National University, Canberra ACT  2601, Australia \and
Department of Nuclear Physics and Accelerator Applications, Research School of Physics, The Australian National University, Canberra ACT  2601, Australia\and
CEA, DAM, DIF, F-91297 Arpajon, France\and
Universit\'e Paris-Saclay, CEA, Laboratoire Mati\`ere sous Conditions Extr\^emes, 91680 Bruy\`eres-Le-Ch\^atel, France}
\date{\today}

\abstract{
Quantum shell effects  stabilising fission fragments with various shapes have been invoked as a factor determining the distribution of nucleons between the fragments at scission. 
Shell effects also induce asymmetric shapes in the nucleus on its way to fission well before the final fragments are (pre)formed. 
These shell effects are studied in fission of $^{180}$Hg, $^{236}$U and $^{256}$Fm with constrained Hartree-Fock-Bogoliubov calculations using the D1S parametrisation of the Gogny interaction. 
Strutinsky shell energy correction  and single-particle energy level density near the Fermi surface are computed. Several neutron and proton shell effects are identified as drivers towards asymmetric fission.
Shell effects are also used to identify the preformation of the fragments in the later stage of fission. 
}
\PACS{
      {PACS-key}{discribing text of that key}   \and
      {PACS-key}{discribing text of that key}
     }

\maketitle







\section{Introduction}

Quantum shell effects are usually associated with a large energy gap of a few MeV in the single-particle energy spectrum near the Fermi level.
Such gaps provide additional binding, and thus more stability, to the system which can be spherical (as in the so-called ``magic'' nuclei) or deformed. 
In the fission process, the nucleus experiences a competition between Coulomb repulsion and nuclear attraction which normally deforms the system up to the formation of two fragments with similar masses.  
On its way to fission, however, the system may encounter shell effects that drive it out of the symmetric fission path, i.e., favouring pear shapes to the nucleus, leading to mass asymmetric fragments~\cite{andreyev2018,schmidt2018}.
The system thus follows one or several fission valleys in a ``potential energy surface'' (PES) representing the potential energy as a function of collective coordinates used to characterise the deformation of the system. 
Its dynamics is then guided by the topology of the PES along with the associated
inertia.

Quantum shells stabilising fission fragments with spherical~\cite{mayer1948,meitner1950}, cigar~\cite{wilkins1976}, and pear~\cite{scamps2018} shapes have been invoked as a factor determining the distribution of nucleons between the fragments at scission. 
These shells, however, are found only in the later stage of the fission process when pre-fragments are already formed, while 
other shell effects are expected in the fissioning nucleus. 
For instance, the decrease in energy of neutron orbitals has been invoked as a mechanism inducing asymmetric shapes near the second saddle point of actinides \cite{gustafsson1971} as well as in neutron deficient mercury region \cite{ichikawa2019}. In addition, a third hyperdeformed minimum, produced by shell effects, is usually found in the fission path of actinides \cite{cwiok1994}.

Shell energy corrections favouring specific elongations and asymmetries on the way to fission can be identified with the Strutinsky method \cite{strutinsky1967,strutinsky1968}. 
Naturally, the shell-energy correction and single-particle level density are intrinsically related as such corrections are the result of energy gaps appearing in the single-particle energy spectra~\cite{brack1972}.
Deformed shell effects on the way to fission can then be investigated through their effect on shell energy correction, or directly from single-particle energy level density near the Fermi surface. 

In this work, shell effects are studied in fission of $^{180}$Hg, $^{236}$U and $^{256}$Fm. These nuclei are  
located in different mass regions of interest for nuclear fission studies~\cite{andreyev2018,schmidt2018} as they are expected to exhibit different shell effects in the final fragments~\cite{scamps2018,staszczak2009,warda2012,scamps2019,mahata2022}.
PES are obtained with  the constrained Hartree-Fock-Bogoliubov (HFB)
method.
In addition to the total potential energy of the deformed system used to construct the PES, the HFB calculations give access to the single-particle energy levels that are used to compute the shell-energy corrections from the Strutinsky method as well as the single-particle energy level density near the Fermi surface.
The method and numerical details are provided in section~\ref{sec:method}. Resulting PES and shell effects are discussed in section~\ref{sec:results}. Conclusions are drawn in section~\ref{sec:conclusions}.

\section{Theoretical method and numerical details\label{sec:method}}

\subsection{Constrained Hartree-Fock-Bogoliubov calculations} 

Quantities of interest are studied within the HFB
approximation under constraints using the D1S parametrisation \cite{berger1991} of the Gogny nuclear interaction \cite{robledo2018}.
Details of the procedure can be found in Ref. \cite{bernard2020}.
Axial ($z-$axis), time reversal and simplex symmetries are preserved in this study.
As a result of the time reversal symmetry, only systems with an even number of protons and neutrons are computed. 

A good representation of PES is usually obtained by imposing quadrupole $Q_{20}$ (cigar shape) and octupole $Q_{30}$ (pear shape) moments, while leaving free the higher multipole moments in an energy minimisation procedure~\cite{schunck2016,schunck2022}.
PES  are generated by constraining 
the quadrupole moment  
$$Q_{20}=\int d^3r \,\rho(\mathbf{r}) \left[z^2-\frac{1}{2}(x^2+y^2)\right]$$
 and the octupole moment  
$$Q_{30}=\int d^3r \,\rho(\mathbf{r}) \left[z^3-\frac{3}{2}z(x^2+y^2)\right],$$
 where $\rho(\mathbf{r})$ is the density of nucleons. 
Higher multipole moments are not constrained in the PES calculations.
The PES are built by minimising the HFB energy for each 
($Q_{20},Q_{30}$) deformation with a 2~b quadrupole moment step and 
a 2~b$^{3/2}$ octupole moment step (1~barn~$\equiv1$~b~$=10^{-28}$~m$^2$).

The resulting constrained HFB states have no internal excitation. Thus, the present calculations are only valid at low energy. 
Indeed, pairing energy and shell effects are affected by excitation energy and thus must be accounted for in fission studies of warm nuclei \cite{mcdonnell2014,ward2017}.\\
The Fock space is spanned in the harmonic oscillator basis whose
number of major shells is $14$ for fissioning nuclei and fragments (see Appendix A). 
The HFB calculations are performed with Lagrange parameters to constrain the number of protons and neutrons in the system.
These Lagrange parameters are chemical potentials, or equivalently Fermi energies marking the transition between occupied and unoccupied states. 
The HFB generalised Hamiltonian is represented by a  matrix containing the usual Hartree-Fock single-particle Hamiltonian $h$ in its diagonal and the pairing field in the off-diagonal terms. 
In this work, the single-particle energies are defined as the eigenvalues of $h$.
The stability of the results is studied in Appendix A in which the size of the harmonic oscillator basis has been raised to $N=15$ for $^{256}$Fm. Its barrier height has been calculated up to $N=24$.

Discontinuities \cite{schunck2016,dubray2012,regnier2019,zdeb2021}
are often associated with a sudden change of shape in the system. They can be identified through a measure of the integrated difference in the density distributions  between two neighbouring points in the PES \cite{dubray2012}, through an evaluation of overlaps between neighbouring states of the PES \cite{verriere2017}, or through the analysis of unconstrained multipole moments, among which the triaxial quadrupole moment $Q_{22}$ and the hexadecapole moment $Q_{40}$  are expected to be the most important  \cite{zdeb2021}.
 We have used the latter method with $Q_{40}$ only  as for an axial code, $Q_{22}$ is constrained to be zero.
Smoothing these discontinuities is beyond the scope of this work and could be achieved using methods proposed in Refs.~\cite{dubray2012,lau2022}.
 
 It is also possible to find more than one local minimum  for a given set of constraints in the construction of the PES. To avoid this issue, one can perform a retropropagation scheme and keep the lowest minimum \cite{regnier2019}. Another approach, which is the one we adopted, is to keep the system in the fission valley as long as possible (see discussion in \cite{zdeb2021}).

\begin{figure*}
   \includegraphics[width=1\linewidth]{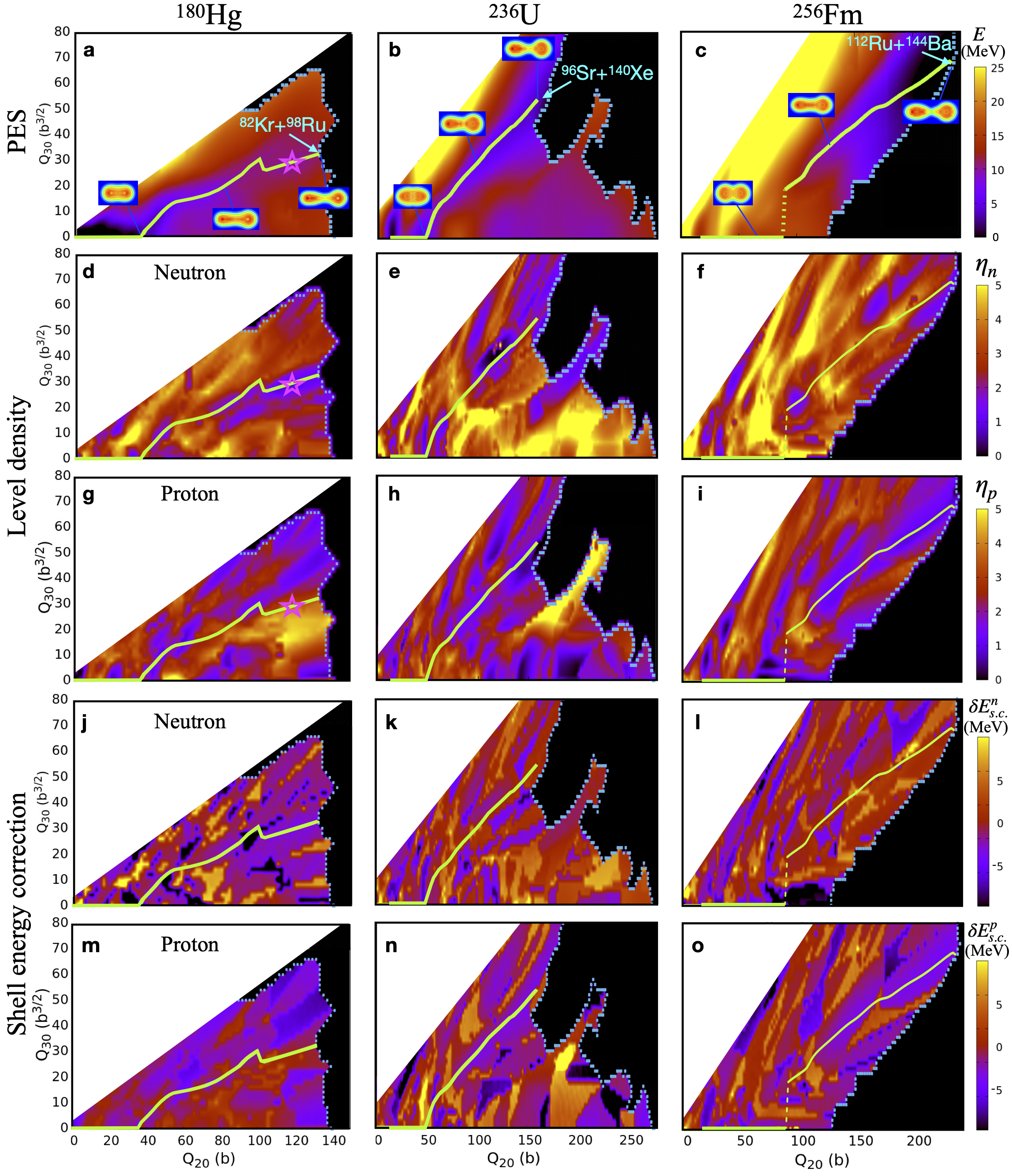}  
\caption{\label{fig:PES}
{(a-c)} The potential energy is plotted as a function of the quadrupole $Q_{20}$ and octupole $Q_{30}$ moments in $^{180}$Hg, $^{236}$U and $^{256}$Fm. Fission paths are represented by  thick green lines. Density contours along the fission path are also shown. 
The most probable fragments produced at scission are indicated by an arrow. 
The dotted segment in ({{c,f,i,l,o}}) indicates a discontinuity between the symmetric and asymmetric valleys (see, e.g., \cite{dubray2012,regnier2019,zdeb2021}) when both valleys have the same energy (in this case, the ridge between both valleys is only $\sim0.5$~MeV). 
The {(d-i)} panels show the neutron ($\eta^n$; {(d-f)}) and proton ($\eta^p$; {(g-i)}) level densities near the Fermi energy, respectively. Shell effects are stronger for darker colours. 
The star {(a,d,g)} indicates the configuration used to produce the single particle level scheme in figure~\ref{fig:occ}.
Finally, neutron (j-l) and proton (m-o) Strutinsky shell energy corrections are plotted from
the HFB single particle energy spectra. 
}
\end{figure*}

The resulting PES of $^{180}$Hg, $^{236}$U and $^{256}$Fm are shown in Fig.~\ref{fig:PES}(a,b,c) and are discussed in Sec.~\ref{sec:results}.

\subsection{Definitions of the fission path, of scission and of the pre-fragments}

In principle the path to fission is the one that leaves the action of the system stationary. 
In this work, what is called ``fission path'' is obtained by assuming that the system follows a minimum energy path on its way to fission. 
Although this approximation is expected to break down when non-adiabatic effects become significant (e.g., near scission) as well as near discontinuities, 
it is sufficient for the purpose of this work that is to investigate the role of shell effects. 
Note that, although collective inertia is calculated (see Appendix C), it is not used in the path calculation. 

Several definitions of scission have been used by different groups (see discussion in \cite{schunck2022}).
Here, it is assumed that scission is achieved  when the spatial density in  the neck becomes lower than $0.08$~fm$^{-3}$
(approximatively half the nuclear saturation density).
Calculations along the fission paths and within the PES are stopped at the last point before
reaching scission when increasing the elongation. This defines the scission point and the
pre-scission line, respectively. 
Although this simple definition of scission could lead to ambiguities in characterising the fragments (e.g., due to their entanglement), it is sufficient in the present work that is devoted to shell effects within the PES up to the scission region, as close as possible to the nascent fragments.

For a given HFB configuration of the fissioning nucleus,  
the spatial density minimum in the neck along
the deformation axis is used to identify left and right
pre-fragments. Integrating the spatial density on the left
and on the right enables us to get pre-fragment mass and 
charge numbers and their individual multipole moment 
deformations 
$$Q_{\lambda0}=\sqrt{\frac{4\pi}{2\lambda+1}}\int d^3r \,\rho(\mathbf{r})r^\lambda Y_{\lambda0}(\theta,\phi)$$ 
for $\lambda\in [2,6]$.
 $Y_{\lambda0}$ are the spherical harmonics and $\theta$ and $\phi$ are the polar and azimuthal angles. 
Particle numbers and these multipole moments 
are then used as constraints to calculate pre-fragments alone.\\

\subsection{Shell energy corrections and single-particle level density near the Fermi surface}

\begin{figure}
   \includegraphics[width=1\linewidth]{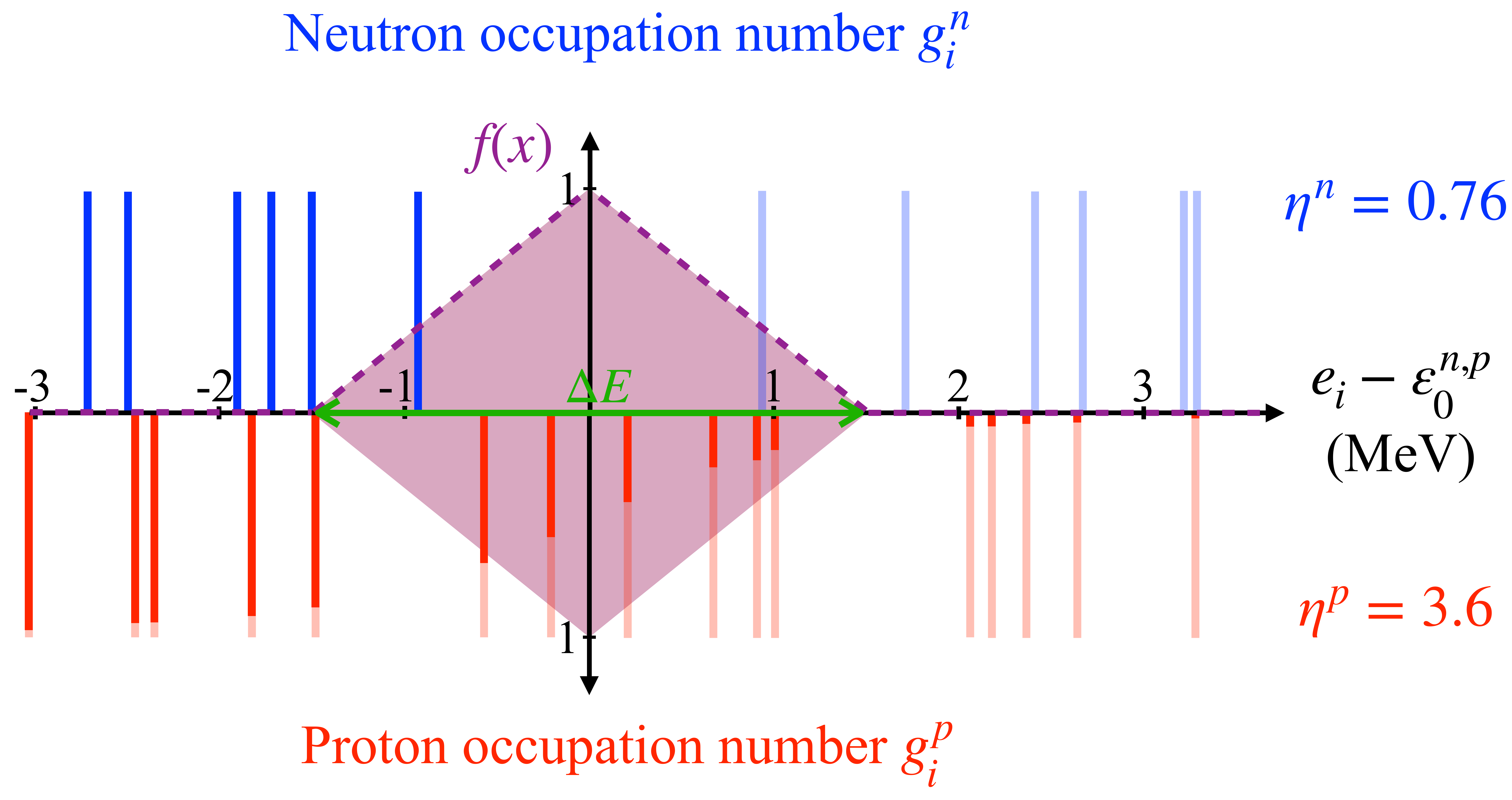}  
\caption{\label{fig:occ}
Occupation numbers $g_i^{n,p}$ (non-transparent part of vertical lines) of nucleon states with energy $e_i$ near the Fermi energy are shown for a deformed state of $^{180}$Hg [corresponding to the star in Figs.~\ref{fig:PES}(a,d,g)]. 
The resulting values $\eta^{n,p}$ of the level densities near Fermi energy are indicated on the right. 
}
\end{figure}

Details of the shell energy corrections method of Strutinsky \cite{strutinsky1967,strutinsky1968} can be found in \cite{ring1980}. The width over 
which the smoothing of the single particle level scheme is performed is 
chosen as $\gamma = 41 A^{-1/3}$, $A$ being the number of neutrons or protons.

The density of energy levels in the system determines important quantities that affect fission mechanisms, including energy dissipation \cite{ward2017}, fragment excitation \cite{albertsson2020}, and shell effects \cite{brack1972}. In the case of shell effects, however, only the level density near the Fermi level is relevant.  
To investigate the evolution of shell effects along the fission path, we introduce an energy level density $$\eta^{n,p}=\sum_i f(e_i-\varepsilon^{n,p}_0)$$ that counts the number of neutron or proton single-particle states in an energy window $\Delta E$ centred in $\varepsilon_0^{n,p}$ that are defined as halfway between the two levels surrounding the neutron and proton Fermi energy, respectively. 
$\Delta E$ should be of the order of typical shell energy gaps so that $\eta^{n,p}$ produces small values in the presence of such a gap.
As the nucleon energy spectra are less compressed in lighter nuclei, we use $\Delta E=3$~MeV in $^{180}$Hg and $2.5$~MeV in the other nuclei. 
To avoid rapid changes in $\eta^{n,p}$ when a nucleon energy level enters or leaves the energy window, we also weight the contribution of the levels by a function $f$ of the nucleon energy $e_i$ that is equal to 1 at $\varepsilon_0^{n,p}$ and linearly decreases to 0 at the edges of the energy window. 
The effect of $\Delta E$ and of the  smoothing function are studied in Appendix~B.

Figure~\ref{fig:occ} 
shows how $\eta^{n}$ and $\eta^p$ are computed for a state of $^{180}$Hg on its fission path [represented by a star in Figs~\ref{fig:PES}(a,d,g)].
In particular, we see that a large energy gap near the neutron Fermi level leads to a small value of $\eta^n$. 
The level densities $\eta^n$ and $\eta^p$ can then be used as an indication of the presence of neutron and proton shell effects, respectively. 
In turn, these shell effects are expected to produce a shell correction energy (as computed with the Strutinsky method) 
and also affect the inertia (and thus the dynamics) of the system as it evolves on the PES \cite{sadhukhan2013}.

\section{Potential energy surfaces and identification of shell effects\label{sec:results}}

Figures~\ref{fig:PES}(a-c) show the PES for $^{180}$Hg, $^{236}$U and $^{256}$Fm.
In each case the fission is mass asymmetric as indicated by the non-zero values of  $Q_{30}$ along the one-dimensional fission path obtained by following the path of local minimum energy in the potential energy surface until scission. 
The predicted  fragment mass asymmetries at the end of the fission path are in good agreement with the centroids of the experimental fragment mass distributions~\cite{andreyev2010,unik1974}.
No  discontinuity is observed along the fission path except a small one at $Q_{20}\simeq100$~b in $^{180}$Hg and a larger one at $Q_{20}\simeq90$~b in $^{256}$Fm. In both cases, they correspond to ``jumps'' in the octupole moment $Q_{30}$  as seen from the PES in Figs.~\ref{fig:PES}(a) and~\ref{fig:PES}(c). We checked that, for the main discontinuity in $^{256}$Fm, $Q_{40}$ evolves smoothly with $Q_{30}$
[See Fig. \ref{Q40andinertia}(c) in Appendix C].
As inertia plays an important role in cold fission, the perturbative masses have been calculated using the adiabatic time-dependent Hartree-Fock-Bogoliubov (ATDHFB)  and 
generator coordinate method (GCM) approximations. They are presented in Appendix~C.

The existence of a third minimum in fission path of actinides
is currently debated \cite{kowal2012,jachimowicz2013,ichikawa2013b,zhao2015}. Here, no third minimum has been found on the $^{236}$U 
asymmetric path. However, $^{256}$Fm experiences a strong first barrier (about $10.5$ MeV) followed by a shallow second minimum at $\sim 52$ b and a shallow third minimum
at $\sim 72$ b. The second and third minima are separated by a  $\sim 1$ MeV barrier.

Figure~\ref{fig:PES} also shows the level densities $\eta^{n,p}$ for neutrons (d-f) and protons (g-i) as well as   Strutinsky shell energy correction from the neutron  (j-l) and proton (m-o) HFB single particle energy spectra for the same three nuclei and ranges of deformations as in the PES. 
Both methods indicate that, in each system, when mass symmetry is broken the fission paths  follow successive regions associated with shell energy correction and, similarly,  small neutron and/or proton level densities. 
We see that several shell effects are at play along the fission path and that both proton and neutron shells may take turn in driving the system towards mass asymmetric fission.

Note also that in the Hartree-Fock-Bogoliubov theory used in this work, static pairing correlations are fully accounted for and as the finite range Gogny interaction is used, no cut off is required for the pairing energy. 
Thus the results on the level density function and the Strutinski shell energy correction based on the PES analysis both include a full treatment of the static pairing
correlations. 
Ambiguities may arise when comparing pre-scission configurations with individual studies of fragments as different Lagrange parameters have to be used to constrain the number of particles in both cases. 
However, our study focuses essentially on the role of shell effects in pre-scission configurations and thus we do not expect our conclusions to be affected by such issues.

The pairing interaction, which occurs essentially near the Fermi level, is attractive 
and increases with single-particle energy level density.
It is then in competition with shell effects that favour low level densities. 
The fact that the fission paths follow regions of low level density shows that 
the dominant effect comes from shell effects rather than pairing energy.

Comparing figures \ref{fig:PES}(d-i) with figures \ref{fig:PES}(j-o), we see that 
the variations of the energy correction display structures over the
2D surfaces that are sometimes similar to the ones in the near Fermi level density 
surfaces 
but also differ in some cases. 
Focussing on the 1D fission paths, both methods predict a dominance of 
proton shell effects at $Q_{20}>70$~b in $^{236}$U [Figs.~\ref{fig:PES}(h,n)] and at 
$Q_{20}>110$~b in $^{256}$Fm [Figs.~\ref{fig:PES}(i,o)]. 
The level density method, however, predicts a region of low level density from neutron shell effects at $Q_{20}\sim100$~b and $Q_{30}\sim20$~b$^{3/2}$ [Fig.~\ref{fig:PES}(f)] that could be responsible for the early asymmetry in $^{256}$Fm, though the latter is not associated with significant shell correction energy [Fig.~\ref{fig:PES}(l)]. 
Similarly, the level density method predicts a neutron gap in $^{236}$U at $100<Q_{20}<130$~b [Fig.~\ref{fig:PES}(e)] that is 
not observed in the shell correction energy [Fig.~\ref{fig:PES}(k)]. 
As for $^{180}$Hg, while both methods agree on the dominance of neutron
shell effects in the later stage of fission [Fig.~\ref{fig:PES}(d,j)], only the level density exhibits structures that are 
clear enough so that we can interpret the discontinuity at $Q_{20}\sim100$~b as due to a change from proton to neutron shell effects.

In the following, we base our analysis of shell effects on the level density plots in Figs.~\ref{fig:PES}(d-i).
In $^{180}$Hg, both proton and neutron shells are present when the system first acquires a non-zero octupole moment $Q_{30}$ [Figs.~\ref{fig:PES}(b,g)]. 
In $^{236}$U, the first asymmetry is induced by neutrons [Fig.~\ref{fig:PES}(e)], while in $^{256}$Fm it is due to protons [Fig.~\ref{fig:PES}(i)].
Indeed, in the latter, the low $\eta^p$ at $Q_{20}\simeq60-90$~b lowers the energy ridge between the symmetric path and a nascent asymmetric one. 
Near scission, the final asymmetry between the fragments of $^{180}$Hg is determined by neutron shells [Fig.~\ref{fig:PES}(d)], while in $^{236}$U and $^{256}$Fm, it is dominated by proton ones [Figs.~\ref{fig:PES}(h,i)]. 
 Strong shell effects are also observed at intermediate configurations along the fission path, such as neutron shells at $Q_{20}\sim120$~b in $^{236}$U [Fig.~\ref{fig:PES}(e)] and 100~b in $^{256}$Fm [Fig.~\ref{fig:PES}(f)].  
  
In the case of $^{236}$U, this confirms that, on the one hand,  neutron shell effects allow the nucleus to leave the symmetric path to find a lower (asymmetric) second barrier \cite{gustafsson1971,brack1972}, and, on the other hand, proton shell effects are present near scission \cite{scamps2018}. We also see that these two mechanisms do not provide a complete picture. Indeed, in $^{236}$U, we observe several additional proton and neutron shell effects on the asymmetric fission path [Figs.~\ref{fig:PES}(e,h)], showing that the connection between the first and last shell effects is not trivial. 
This observation is in contradiction with an earlier interpretation (in the case of $^{232}$Th) that a single shell effect due to a partially formed spherical Sn-like fragment is responsible for lowering the level density from the asymmetric saddle to scission~\cite{moller2009}. 

\begin{figure*}
   \includegraphics[width=1\linewidth]{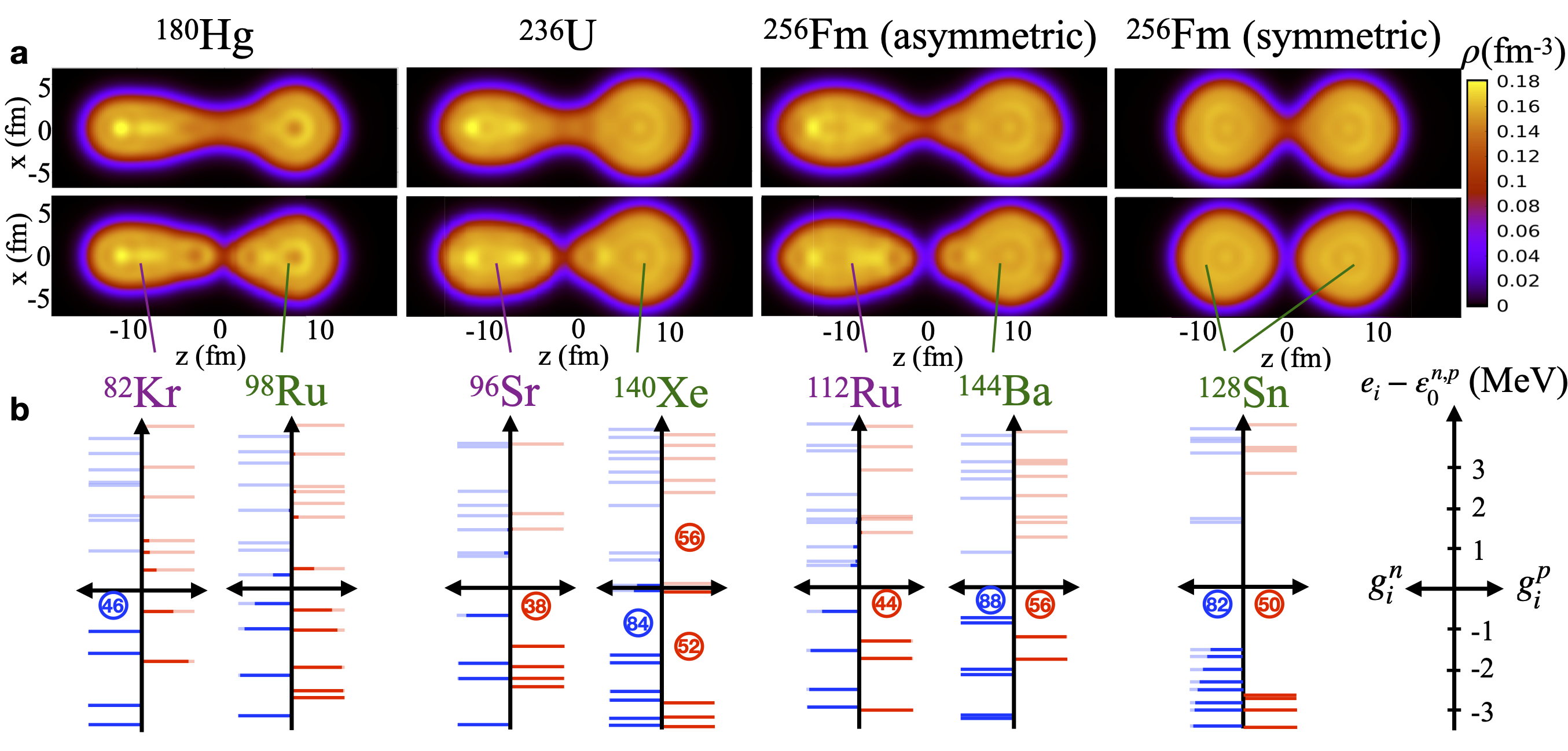}  
\caption{\label{fig:frag}
{(a)} The density distributions of the nuclei on their fission path just before scission are shown in the top panels. Below each panel are shown the density distributions for the ``closest'' fragments with even numbers of protons and neutrons. 
{(b)} The resulting proton and neutron energy levels $e_i$ and occupation numbers $g_i^{n,p}$  (non-transparent part of horizontal lines) near the Fermi energy  are shown for each fragment. The numbers in the energy gaps correspond to the number of particles that can occupy the single-particle states below the gap. 
}
\end{figure*}

Note that shell effects are not limited to the fission path (which is only the most probable one) and may affect other regions of the PES which could influence fragment mass and charge distributions. 
For instance, by comparing Figs.~\ref{fig:PES}(b),  \ref{fig:PES}(e) and \ref{fig:PES}(h) we see that in $^{236}$U, proton shell effects produce another fission valley which is symmetric ($Q_{30}\simeq0$), while the extension of the PES around $Q_{20}\sim200-220$~b and $Q_{30}\sim30-50$~b$^{3/2}$ is clearly due to neutron shell effects. In addition, a shallow symmetric valley in $^{256}$Fm [Fig.~\ref{fig:PES}(c)], produced by spherical shell effects in the fragments leading to low proton level density [Fig.~\ref{fig:PES}(i)], is  observed (see also Fig.~\ref{fig:frag}).

In principle, both the level density and the shell energy correction methods are able to predict at which stage the final fragments are pre-formed in the system, while still interacting via the strong force through the neck. 
Fragments can be considered as being pre-formed when the nucleons occupy the states that are necessary to build their final microscopic structure, even though the shapes of the pre-formed fragments may not be the final ones. 
If a shell energy gap, characterised by a low level density at the Fermi level, is present and extends up to scission, we can then consider the fragments to be pre-formed in this region of the PES. Indeed, the presence of the gap means, in principle, that no state is crossing the Fermi level, thus all the ``required'' states for the final fragments are already occupied~\cite{simenel2014a} (at least for the specie of nucleons -- neutrons or protons -- that exhibits the gap). 
For example, the neutron shell gap that constrains the fission path of $^{180}$Hg near scission appears at $Q_{20}\simeq100$~b, at which point the fragments can be considered to be pre-formed. 

The methods of identifying shell effects from shell energy corrections and from the level density provide information on the entire system, such as which of proton and/or neutron shells are the main drivers towards fission.
Once the fragments are pre-formed, however, they have their own microscopic structure which we can use to investigate which of the fragments exhibits shell effects fixing the final mass asymmetry~\cite{scamps2018,scamps2019,mahata2022}. 
A similar analysis is presented in figure~\ref{fig:frag}.
The fragments are identified by integrating the proton and neutron densities on each side of the neck, at the most probable scission configuration (i.e., at the end of the fission path). 
This only gives an average number of protons and neutrons. 
However,  the shell structure of a fragment is expected to show only small variations by adding or removing one nucleon.
This allows us to determine the proton and neutron levels of the fragments from microscopic calculations of nuclei with even numbers of protons and neutrons. 
Nevertheless, knowing the numbers of protons and neutrons is not sufficient as the shell structure is affected by the deformation of the nuclei. 
It is therefore important to impose deformations in the calculations of the individual nuclei which are close to those of the fragments  observed in the total system at scission. 
The fragments are then computed individually (without interaction between them), with constraints on their multipole moments which are optimised to reproduce the scission configuration. 
This allows us to compute the single-particle spectra for each fragment. 
In figure~\ref{fig:frag}, this is done by adjusting the multipole moments $Q_{\lambda0}$ of the simulated fragments with $\lambda=2-6$.

Figure~\ref{fig:frag} shows that the low level densities in compound nuclei identified at scission are associated with shell gaps at the Fermi level in the fragments.
A neutron shell gap at $N=46$ is present in the light fragment ($^{82}$Kr) of $^{180}$Hg, confirming that neutron shells drive this system near scission. 
A similar situation occurs with proton shells in both fragments of $^{236}$U and $^{256}$Fm with, 
in addition, the presence of neutron shell gaps  in the heavy fragments which could also impact the fission process. 
The fact that shell effects can be found in both fragments does not necessarily mean, however, that both fragments have an equal influence in the final asymmetric configuration. 
Indeed, shell effects in the fission fragments could either be a cause or a consequence of the most likely state at scission.  
As mentioned earlier, shell effects provide an additional stability to the system, thus favouring a specific shape in which such shell effects can be found. 
At scission, each fragment can have a different deformation, e.g., one fragment can be compact and the other one elongated, as  in the systems we have studied [see Fig.~\ref{fig:frag}(a)]. 
A leading shell effect
could stabilise the shape of one fragment whereas 
the complementary fragment minimises the energy while preserving the compound nucleus elongation and asymmetry. 
Under this overall
deformation constraint
it is then possible that one fragment is responsible for the configuration at scission, thus driving the final asymmetry, while the other fragment ``just'' finds a deformed shell effect. 

In actinide nuclei such as $^{236}$U and $^{256}$Fm, experimental~\cite{unik1974,schmidt2000,bockstiegel2008,chatillon2019,vermeulen2020} systematic studies indicate an influence of proton shell effects on the heavy fission fragment properties. 
One possible origin of these quantum shell effects is octupole correlations favouring pear shapes in the fragments~\cite{scamps2018}.
In lighter nuclei in the lead region (such as $^{180}$Hg), the origin of the shell effects driving the final asymmetry at scission is still debated~\cite{warda2012,scamps2019,mahata2022,prasad2020,swinton2020b}. 
Future studies 
of the role of shell effects in the fission paths should provide a deeper insight into this region.
 More generally, signatures indicating which shell effect (if any) is the main driver could be searched for in the evolution of the system in the fragment pre-formation region. Indeed, the same technique presented in Fig.~\ref{fig:frag} to analyse the fragments at scission can be used to investigate less elongated configurations, prior to scission. 
 
\section{Conclusions\label{sec:conclusions}}

The role of shell effects in fission has been investigated with the shell correction method of Strutinsky together with the single-particle energy level density near the Fermi surface.
Both methods show that the shape evolution of the fissioning nucleus is determined by several quantum shells that  are driving the nucleus towards mass asymmetric fission well before the final fragments are (pre)formed. 
In the three nuclides we investigated ($^{180}$Hg, $^{236}$U and $^{256}$Fm), at least 1 to 2 neutron and 2 to 3 proton shell effects can be identified on the way to fission.
Several (but not all) of these shell effects lead to significant shell energy corrections. 
The appearance of shell effects attributed to the fragments can also be used to sign their pre-formation. 
These pre-fragments at the end of the fission path are identified as $^{180}$Hg$\rightarrow^{82}$Kr$+^{98}$Ru,
$^{236}$U$\rightarrow^{96}$Sr$+^{140}$Xe, and $^{256}$Fm$\rightarrow^{112}$Ru$+^{144}$Ba. 
The analysis of their single-particle structures indicates that several deformed shell gaps are present in these pre-fragments: $N=46$, $Z=38,44$ with elongated shapes, and $Z=52,56$, $N=84,88$ with compact octupole shapes. 
Although our focus were on asymmetric fission, similar methods could also be used to investigate symmetric modes, e.g., $^{256}$Fm$\rightarrow^{128}$Sn$+^{128}$Sn that exhibits spherical shell gaps in the fragments. 
Furthermore, it would be interesting to study the microscopic origin of the new compact symmetric fission mode in light thorium isotopes recently observed experimentally~\cite{chatillon2020}. 

Recent studies with time-dependent extensions to the mean-field approach that was used here (see \cite{simenel2018} for a review)
 have also shown the importance of shell effects in quasi-fission reaction \cite{godbey2019,simenel2021,li2022}. 
 Combined with shell effect identification methods as used in this work, such time-dependent approaches could be used to investigate the interplay between energy dissipation and shell effects that would impact the energy released in nuclear fission. 
A complete theoretical description of the fission process remains an important challenge which  requires further development of fully microscopic approaches to tackle the complexity of the  fission mechanism~\cite{bender2020}, as illustrated by the present  observation that many proton and neutron shells are at play in each nuclides.

\section*{Acknowledgments}
We are grateful to D. J. Hinde for his continuous support during this work.
We thank L.M. Robledo for providing his code. 
This work has been supported by the Australian Research Council under Grant No. DP190100256. 
This work was supported by computational resources provided by the Australian Government through the National Computational Infrastructure (NCI) under the ANU Merit Allocation Scheme.

\section*{APPENDIX A: Convergence with the number of harmonic oscillator shells.}
\begin{figure*}[!h]
\begin{center}
   \includegraphics[width=1\linewidth]{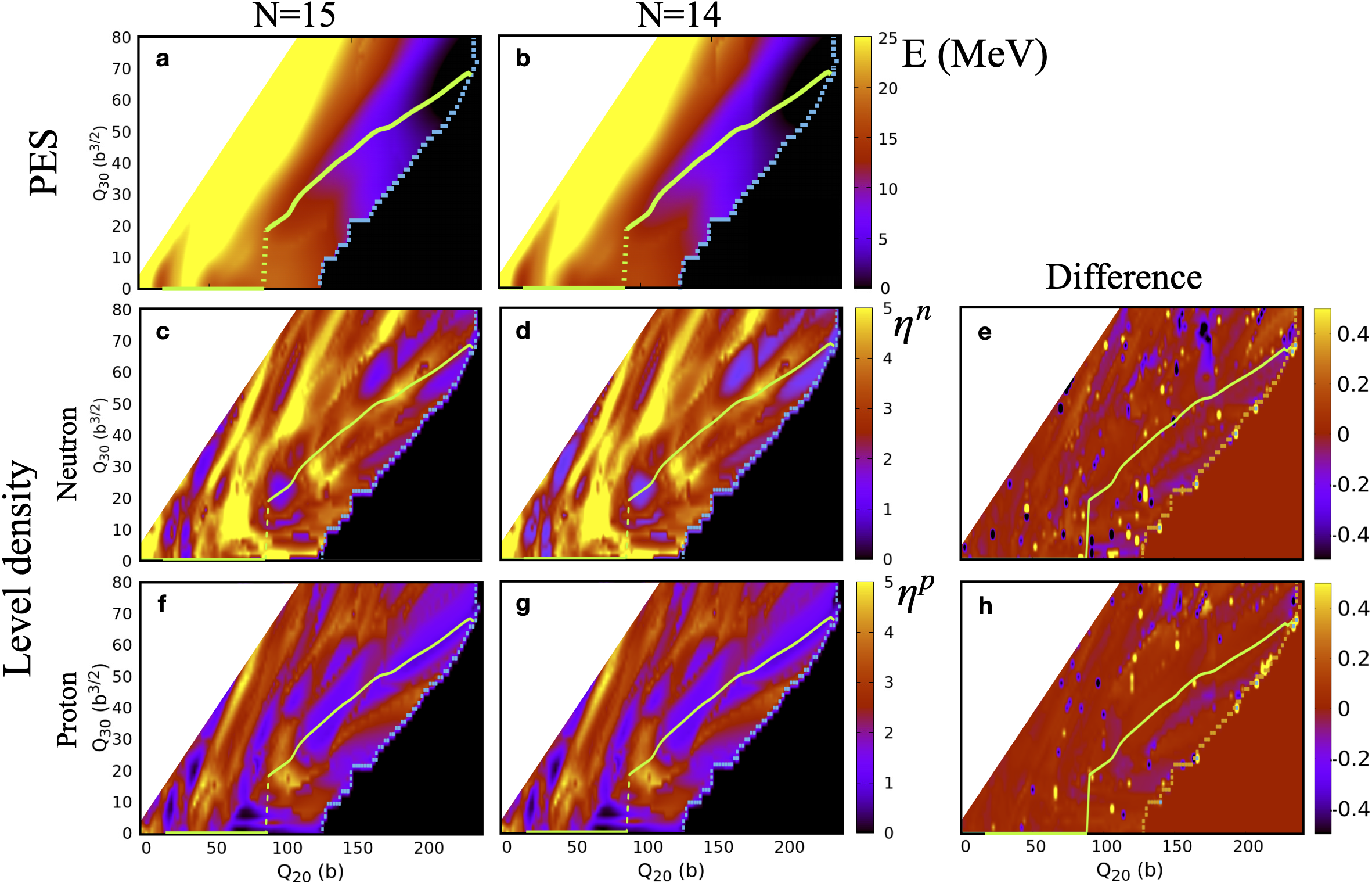}  
\end{center}
\caption{$^{256}$Fm PES for two difference basis sizes: (a) number of major shells $N=15$ and (b) $N=14$. Corresponding level density functions are presented for neutrons
(c,d) and protons (f,g). Their respective differences are depicted in panels (e) and (h),
respectively.
} 
\label{HOdiff}
\end{figure*}

To test the convergence of our calculations with the number of harmonic oscillator shells ($N=14$ in this work), the PES and level densities have been computed for $N=15$ shells in the case of $^{256}$Fm. 
Results are presented in Fig.~\ref{HOdiff}(a) for the PES and
Figs.~\ref{HOdiff}(c,f) for level densities.
The results shown in Fig.~\ref{fig:PES}(c,f,i) with $N = 14$ are
reported in Figs.~\ref{HOdiff}(b,d,g) for comparison. The differences between level densities computed with $N = 14$ and $N = 15$ are small as shown in Figs.~\ref{HOdiff}(e,h), which
is a good indicator of the convergence of our calculations with the number of oscillator shells. 
In particular, they do not exceed 0.5 and usually remain smaller than 0.2 except in small areas of the PES.

Further convergence tests up to $N = 16$ have
been made at specific configurations for $^{256}$Fm, namely the ground state and the barrier height, giving $E_{GS} =-1891.998$ MeV ($N =14$), $-1892.412$ MeV 
(N = 15), and -1892.710 MeV (N = 16) for ground-state energy and $B = 11.220$ MeV ($N = 14$), $11.248$ MeV ($N = 15$) and $11.265$ MeV ($N = 16$) for the
barrier $B = E_{saddle}-E_{GS}$. When increasing further the number of major shells these numbers are found to be stable with: $B = 11.236$ MeV ($N = 20$) and $B = 11.175$ MeV ($N = 24$).

\section*{APPENDIX B: Comparison between several definitions of the density level near the Fermi energy}
 
\begin{figure*}[!h]
\begin{center}
   \includegraphics[width=1\linewidth]{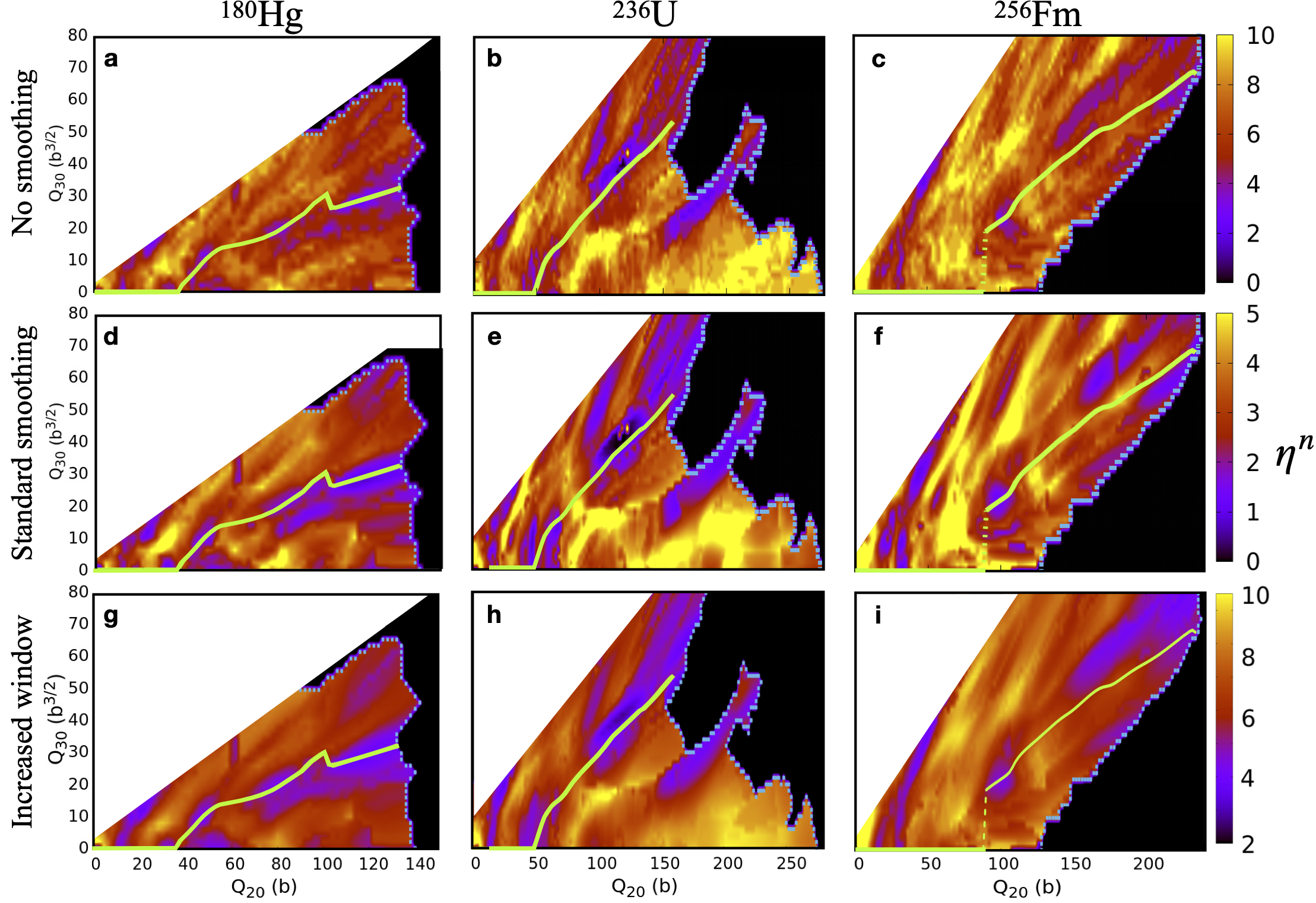}  
\end{center}
\caption{The neutron level density near the Fermi energy is computed without smoothing function within the energy windows $\Delta E=3$ MeV in (a) $^{180}$Hg and $\Delta E=2.5$ MeV in (b) $^{236}$U and (c) $^{256}$Fm.
Panels (d-f) are the same as in Fig.~\ref{fig:PES}. 
Level densities in panels (g-i) are obtained with the smoothing function applied on energy windows increased by a factor 2.
\label{etan}}
\end{figure*}

\begin{figure*}[!h]
\begin{center}
   \includegraphics[width=1\linewidth]{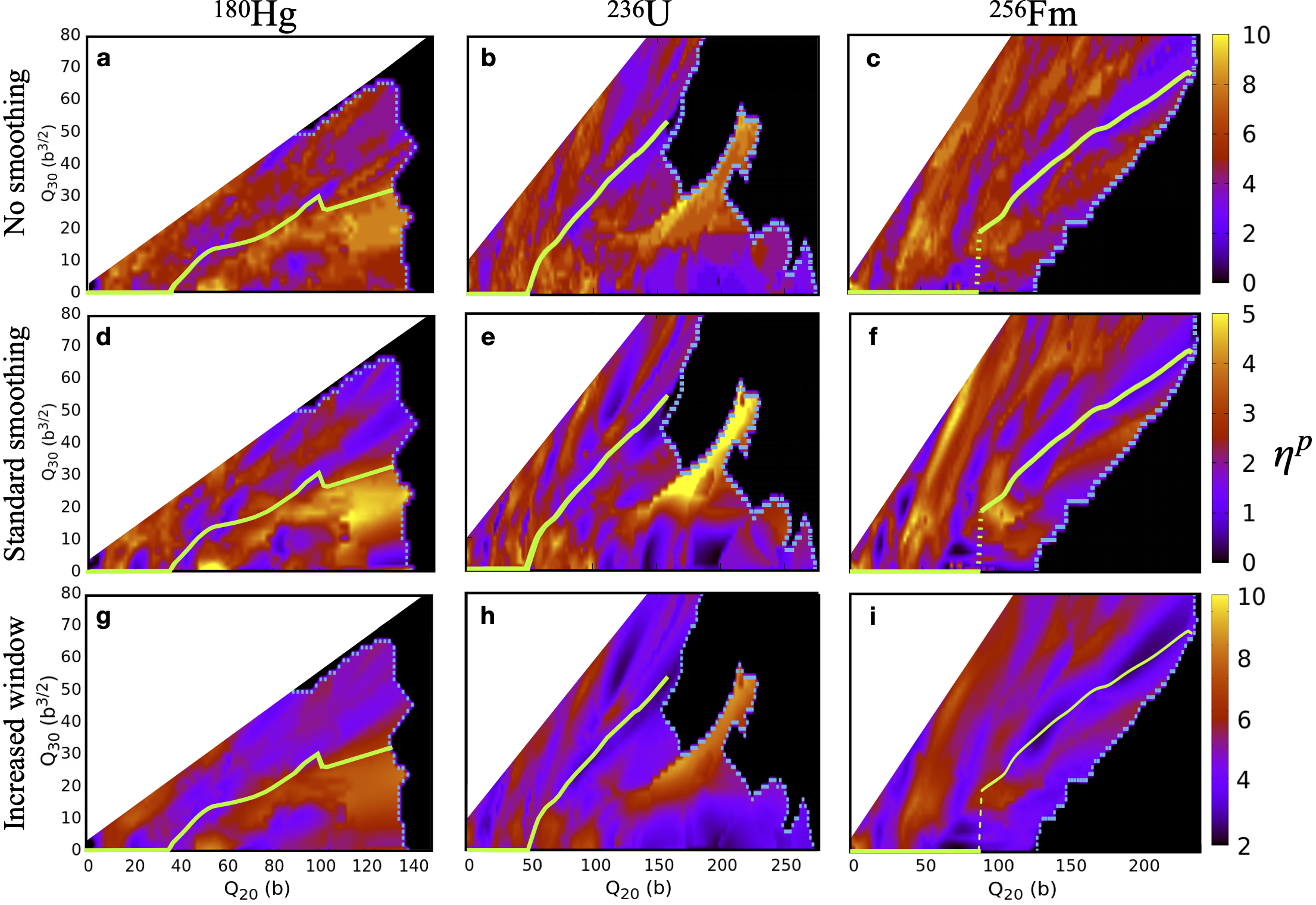}  
\end{center}
\caption{Same as Fig.~\ref{etan} for protons.} 
\label{etap}
\end{figure*}

The energy level density near the Fermi level is defined as $\eta^{n,p}=\sum_i f(e_i-\varepsilon^{n,p}_0)$. 
The effect of the smoothing function $f$ (defined in Fig.~\ref{fig:occ}) and of the energy window $\Delta E$ in which it is non-zero are illustrated in Figs.~\ref{etan} and \ref{etap} for neutrons and protons, respectively. 
The smoothing is removed in Figs.~\ref{etan}(a-c) and Figs.~\ref{etap}(a-c), while the energy window is doubled in Figs.~\ref{etan}(g-i) and Figs.~\ref{etap}(g-i). Results with the standard smoothing defined in the manuscript are reproduced in Figs.~\ref{etan}(d-f) and Figs.~\ref{etap}(d-f)
 for comparison. 
 The same shell effects are visible with these three definitions, and thus our conclusions are not sensitive to (reasonable) variations of the energy window and the smoothing function.   

\section*{APPENDIX C: Properties of the 1D fission paths \label{appc}}

\begin{figure*}[!h]
\begin{center}
   \includegraphics[width=1\linewidth]{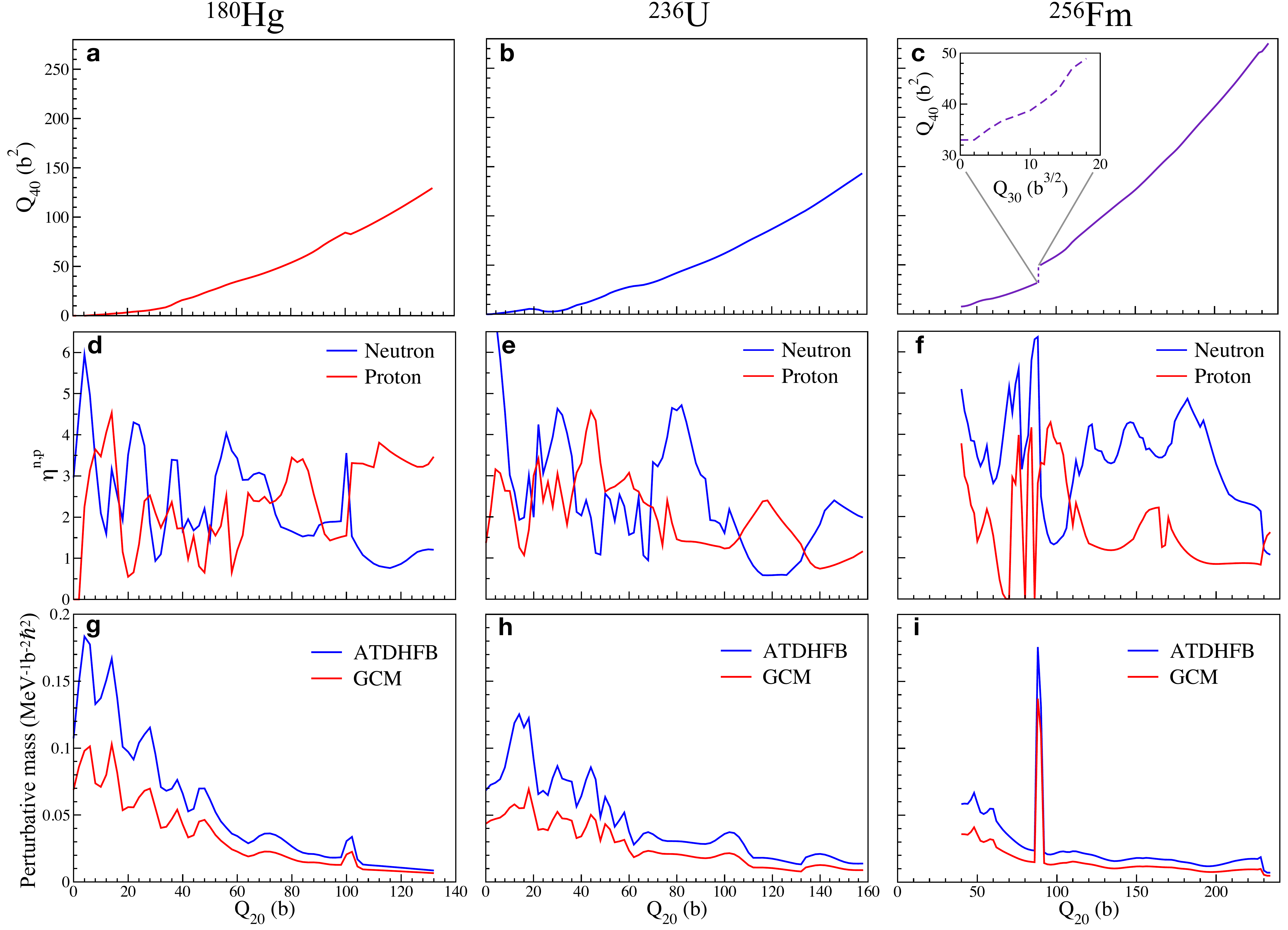}  
\end{center}
\caption{Hexadecapole moments (a-c), smoothed level densities (d-f) and perturbative masses (g-i) as a function of quadrupole moment along the 1D asymmetric fission paths of the three nuclei. The inset of panel (c) shows the evolution of the hexadecapole moment as a function of the octupole moment in the region of the discontinuity.  See Appendix C for details.} 
\label{Q40andinertia}
\end{figure*}

Characteristics of the 1D asymmetric paths are presented  in Fig.~\ref{Q40andinertia}. 
The evolution of the hexadecapole moment $Q_{40}$ along the fission path is plotted for (a) $^{180}$Hg, (b) $^{236}$U, and (c) $^{256}$Fm as a function of the
quadrupole moment $Q_{20}$. No discontinuity is observed for this degree of freedom except a small one at $Q_{20} \simeq 100$ b in $^{180}$Hg and a larger one
at $Q_{20} \simeq 90$ b in $^{256}$Fm. In both cases, they correspond to “jumps” in the octupole moment $Q_{30}$ as seen from the PES in Figs.~\ref{fig:PES}(a,c). The inset in Fig.~\ref{Q40andinertia}(c) shows that, for the main discontinuity in $^{256}$Fm, $Q_{40}$ evolves smoothly with $Q_{30}$, indicating that no higher
multipole moments are necessary to explain this discontinuity.
The evolution of the neutron and proton level densities $\eta^{n,p}$ along the 1D paths extracted from Figs.~\ref{fig:PES}(d-i) are plotted in Figs.~\ref{Q40andinertia}(d,e,f) for the same
systems. 

The perturbative mass parameters accounting for the inertia of the systems are shown in Figs.~\ref{Q40andinertia}(g,h,i). They have been computed
 via two standard methods: the adiabatic time-dependent Hartree-Fock-Bogoliubov (ATDHFB) approach and the
generator coordinate method (GCM) (See Ref. \cite{sadhukhan2013} for details). The large peak observed at $\sim 90$ b in $^{256}$Fm [Fig.~\ref{Q40andinertia}(i)] and the smaller one at $\sim 100$ b in $^{180}$Hg [Fig.~\ref{Q40andinertia}(a)] are associated with
the “jumps” in the PES [Figs.~\ref{fig:PES}(c,a)] and in $Q_{40}$ [Figs.~\ref{Q40andinertia}(c,a)]. 

Level crossing at the Fermi surface are known to produce peaks in
the mass parameters, which is why we see more structure in the inertia evolution at small $Q_{20}$. Naturally, the level density near the Fermi level
is lowered by such crossing, while in regions of the PES affected by strong shell effects, no such crossing is expected. As a result, we expect
correlations between mass parameters and level density at the Fermi surface. When shell effects are strong, such as in the $^{236}$U neutrons at
$Q_{20} \sim 110-130$ b in the fission path, we indeed see a peak in the mass parameter on each side of this region.

\bibliographystyle{epj}
\bibliography{VU_bibtex_master.bib}
\end{document}